\documentclass[page-classic]{epl2}
\usepackage{graphicx, epsfig, rotating}
\begin{document}

\title{Role of Ni-Mn hybridization in magnetism of martensitic state of Ni-Mn-In shape memory alloys}
\shorttitle{Role of Ni-Mn hybridization in Ni-Mn-In shape memory alloys}
\author{K. R. Priolkar\inst{1} \and D. N. Lobo\inst{1} \and P. A. Bhobe\inst{2} \and S. Emura\inst{3} and A. K. Nigam\inst{4}}
\shortauthor{K. R. Priolkar \etal}
\institute{\inst{1} Department of Physics, Goa University, Taleigao Plateau, Goa 403 206 India\\
\inst{2}Institute for Solid State Physics, The University of Tokyo, Kashiwa, Chiba 277-8581, Japan\\
\inst{3}Institute of Scientific and Industrial Research, Osaka University, 8-1 Mihogaoka, Ibaraki, Osaka 567-0047
Japan\\ \inst{4} Tata Institute of Fundamental Research, Homi Bhabha Road, Colaba, Mumbai 400 005 India}

 {Other ferromagnetic metals and alloys}  {Absorption edges, xray}

\abstract {Extended X-ray Absorption Fine Structure (EXAFS) studies on Ni$_{50}$Mn$_{25+x}$In$_{25-x}$ have been
carried out at Ni and Mn K edge as a function of temperature. Thermal evolution of nearest neighbor Ni-Mn and
Mn-Mn bond distances in the martensitic phase give a clear evidence of a close relation between structural and magnetic degrees of freedom in these alloys. In particular, the study highlights the role of Ni$3d$ - Mn$3d$ hybridization in
the magnetism of martensitic phase of these alloys.}

\maketitle

The presence of competing magnetic interactions in the martensitic state has been demonstrated by presence of exchange bias phenomena in Ni$_{50}$Mn$_{25+x}$Z$_{25-x}$ (Z = In, Sn, Sb) alloys \cite{khan2,li,pathak,pab2}. In contrast to Ni$_2$MnGa based alloys \cite{web} a drop in magnetization is observed in these alloys upon structural transformation and has been attributed to enhancement of antiferromagnetic interactions in the martensitic phase \cite{kren1,brown,kano,khan,kren2,buch1,buch2}. Transformation from high temperature austenitic to low temperature martensitic phase with zero or greatly reduced magnetization in Ni$_{50}$Mn$_{25+x}$Z$_{25-x}$ gives rise to number of interesting effects such has inverse magnetocaloric effect, \cite{nat-mat,han} giant strain due to field induced reverse martensitic transformation \cite{kai,koy} and giant magneto resistivity \cite{koy2,yu}.

Uderstanding the magnetism of martensitic state of alloys with excess Mn has been a challenge. While in a X$_2$YZ type Heusler alloy, ferromagnetism arises due to the Ruderman-Kittel-Kasuya-Yoshida (RKKY) exchange interaction between Mn atoms occupying the Y sublattice (Mn(Y)), the origin of antiferromagnetic interactions in Heusler alloys with excess Mn have been a matter of debate. On one hand the antiferromagnetism is believed to be a result of RKKY interactions between Mn atoms in the Z sublattice (Mn(Z)) and those in the Y sublattice \cite{buch2}. The second study proposes  superexchange interactions arising due to hybridization between Ni$3d$ and Mn$3d$ bands as the origin of antiferromagnetic interactions \cite{sasi1,sasi2}. Magnetic state in these alloys is then a result of competition between the ferromagnetic and antiferromagnetic interactions leading to an observation of exchange bias effect. A proper understanding of the mechanism of magnetism of martensitic phase is still to be achieved. Experimental studies have related the change in magnetic ordering of the martensitic phase to the changes in bond distances between the constituent atoms in these alloys \cite{kai-mos}. X-ray Absorption and X-ray magnetic circular dichroism studies on Ni$_2$MnGa type alloys have highlighted the role of hybridization between Ni and Ga atoms in martensitic transformation and the intimate connection between structural and magnetic transformations \cite{jakob} indicating the importance of local structure in influencing the magnetic interactions in these alloys. 

Furthermore, in alloys with excess Mn, EXAFS studies have shown the presence of local structural distortions even in the austenitic phase and that there is a charge transfer from Ni $3d$ band to Mn $3d$ band \cite{pab3,pab4}. Therefore a better understanding of temperature variation of near neighbour bond distances in the martensitic phase of such Heusler alloys is required. Such a study will help to elucidate the magnetic interactions at play in the martensitic phase. It is with this aim that EXAFS studies at the Ni and Mn K edges as a function of temperature in the temperature range 30K $\le$ T $\le$ 300K in case of two alloys of type Ni$_{50}$Mn$_{25+x}$In$_{25-x}$ with $x$ = 0 and 10 have been carried out. Here, $x$ = 0 alloy has a stable crystal structure and serves as a reference point while the other alloy undergoes a martensitic transformation at about 265K. The results obtained  clearly point to the role of nearest neighbor Ni-Mn bond distance in the magnetism of these alloys in the martensitic phase.

The alloys of composition Ni$_{50}$Mn$_{25}$In$_{25}$ (IN25) and Ni$_{50}$Mn$_{35}$In$_{15}$ (IN15) were prepared by arc melting the constituents taken in stoichiometric ratio. The buttons obtained after several melts were sealed in quartz tube and annealed for 48 hours at 750$^\circ$C and later quenched in ice cold water. The ingots were then cut in suitable sizes for further measurements. The samples were characterized by X-ray diffraction (XRD) for phase purity and structure. Energy dispersive analysis by X-rays (EDAX) was employed to check the composition ratios. The constituent atomic ratios obtained for the two samples were Ni = 50.5, Mn = 25.3, In = 24.2 for IN25 and Ni = 51.3, Mn = 33.7 and In = 15.0 for IN15 sample. Magnetization as a function of temperature and field was recorded using a MPMS SQUID magnetometer in the temperature interval 5 K - 350 K and field of 100 Oe. For this measurements the sample were cooled in zero applied field from room temperature to 5K, upon which the field was applied and magnetization was recorded during warming (ZFC), cooling (FC) and subsequent warming (FH) without removing the applied field. EXAFS measurements were performed at  the Photon Factory using beamline 12C at the Mn and Ni K-edges at six temperatures in the interval 30K to 300K. Typically, for EXAFS, the scan was recorded from -300eV to 1000eV with respect to the edge energy in steps of 5eV in the pre-edge region (-300eV to -50e), 1eV in the edge region (-50eV to 100eV) and 2eV in the rest of the region (100eV to 1000eV). Absorbers were prepared by sandwiching appropriate number of sample coated scotch tape layers such that the edge jump $(\Delta\mu(E)) \le 1$. Data analysis was carried out using IFEFFIT \cite{newville} in ATHENA and ARTEMIS programs \cite{ravel}. Here theoretical fitting standards were computed with FEFF6 \cite{ravel2,zabinsky}. The data in the k range of (2 - 12) \AA$^{-1}$ and R range 1 to 3 \AA \ for Ni EXAFS and 1 to 5 \AA \ for Mn EXAFS was used for analysis.

\begin{figure}
\onefigure[width=\columnwidth]{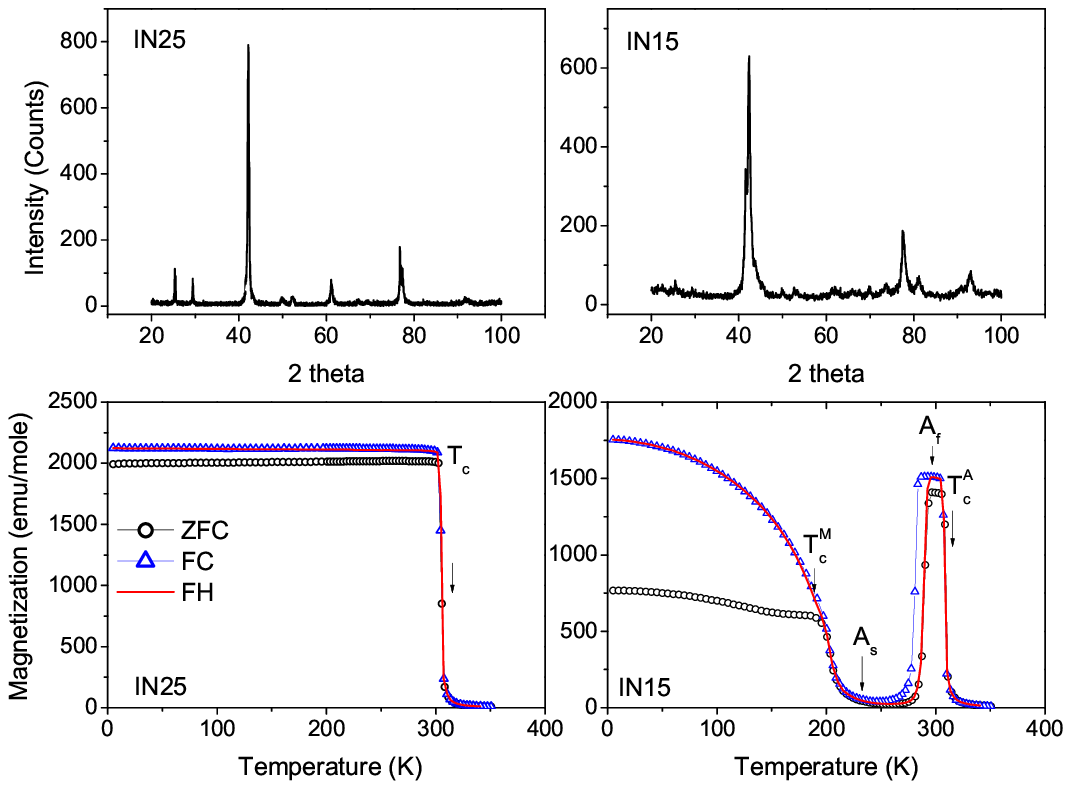}
  \caption{X-ray diffraction patterns (upper panel) and magnetization data (lower panel) recorded in a field of
  100 Oe during ZFC, FC and FH cycles for Ni$_{50}$Mn$_{25}$In$_{25}$ (IN25) and  Ni$_{50}$Mn$_{35}$In$_{15}$ (IN15).}
  \label{fig1}
\end{figure}

X-ray diffraction patterns recorded at room temperature (293K) and magnetization measured during ZFC, FC and FH cycles is plotted in Figure \ref{fig1}. The diffraction pattern for IN25 shows super structure reflections of the $L2_1$ phase indicating a ordered structure with a lattice constant $a$ = 6.067\AA. The corresponding magnetization data also shows a clear ferromagnetic transition at T$_c$ = 306K. In the case of IN15 the XRD pattern shows extra reflections which can be indexed to orthorhombic martensitic phase in addition to those belonging to the cubic austenitic phase. This can be related to the proximity of room temperature to martensitic transformation temperature in this sample. The magnetization measurements indicate the austenitic finish ($A_f$) temperature for this sample to be 290K and the martensitic start ($M_s$) temperature to be 240K. Magnetization curve also shows that the sample undergoes paramagnetic to ferromagnetic transition in the austenitic phase at T$_c^A$ = 310K and another magnetic ordering transition in the martensitic phase at T$_c^M$ = 200K. Therefore it is quite clear that while IN25 is structurally and magnetically stable alloy, IN15 undergoes structural (martensitic) as well as a change in its magentic ordering at low temperatures. A comparative study of local structure around the magnetic ions will help in understanding the role of near neighbour interactions in the magnetism of these two alloys. 

\begin{figure}
\onefigure[width=\columnwidth]{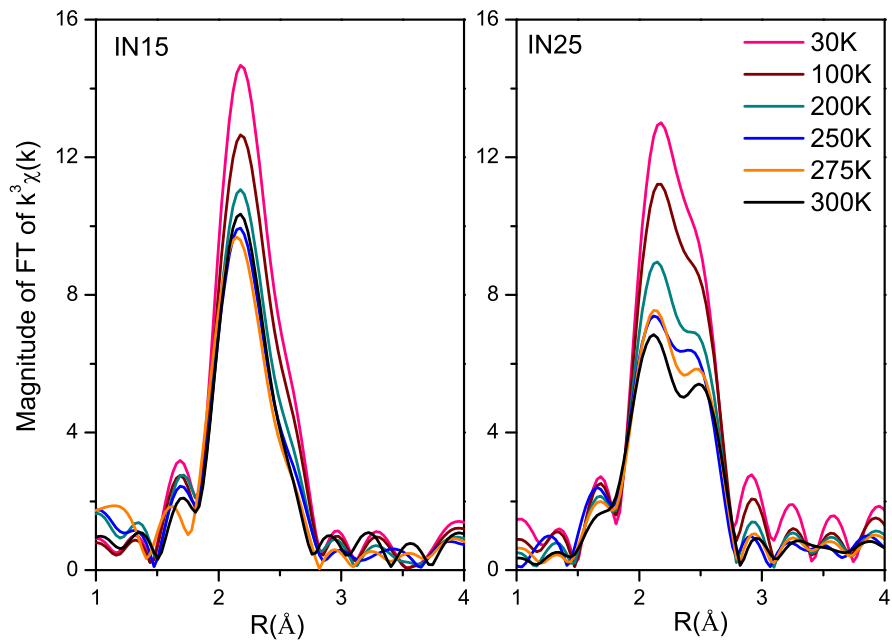}
  \caption{Variation of amplitude of FT of Ni K edge EXAFS as a function of temperature in IN25 and IN15.}
  \label{fig2}
\end{figure}

The magnitude of Fourier transform (FT) of Ni K-edge EXAFS signals recorded at different temperatures for both the alloys is plotted in Figure \ref{fig2}. With the lowering of temperature, the amplitude of the EXAFS oscillations should increase due to decrease in Debye Waller contribution to $\sigma^2$. This is clearly evident in the case of IN25 in Figure \ref{fig2} wherein the peak in the FT spectra of EXAFS becomes sharper with decrease in temperature. Whereas, for IN15, the FT amplitude shows a sudden decrease in the temperature interval 300K $<$ T $\le$ 250K. This being the region of martensitic transformation, a decrease in FT peak amplitude can be related to structural disorder present in the alloy. The second feature visible from Figure \ref{fig2} is the change in the position in the temperature interval 300K to 250K of the first peak in the FT spectra of IN15 alloy.  The position of the peak in FT spectra corresponds to bond distance between the absorbing and scattering atom and therefore a change in its position indicates a variation in bond distance.

\begin{figure}
 \onefigure[width=\columnwidth]{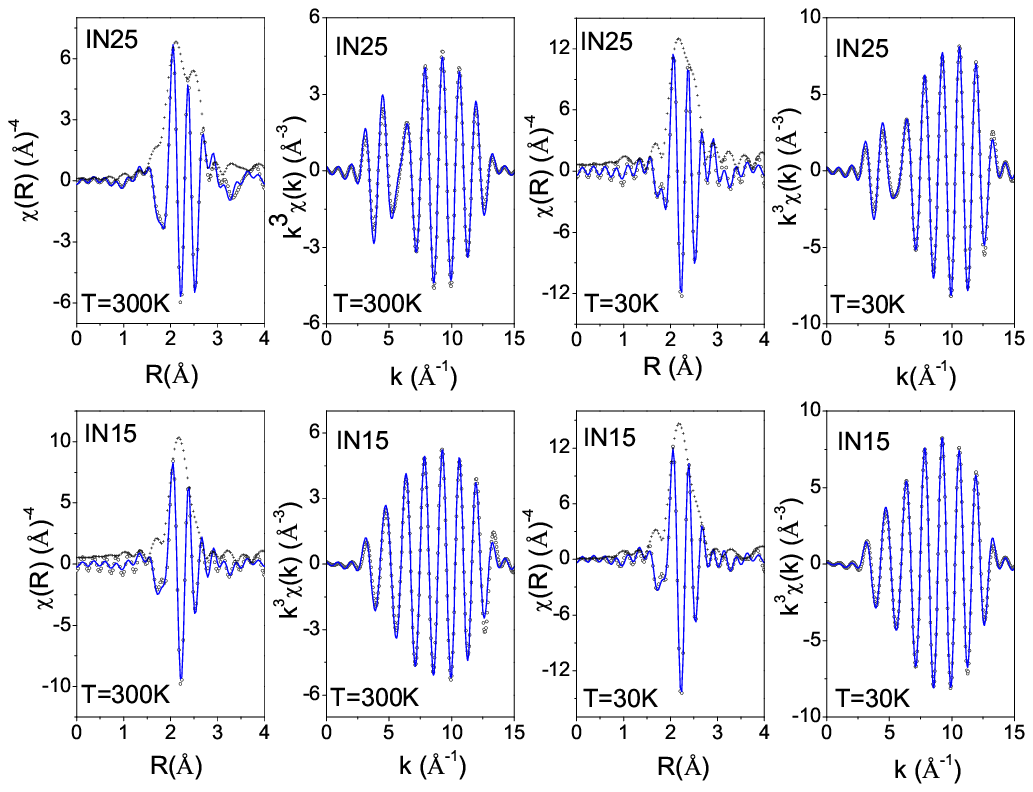}
  \caption{Magnitude and real component of FT of EXAFS spectra in R space and real component in the back transformed $k$ space for
Ni K-edge in IN25 and IN15 at 300K and 30K. The fitting to the data are shown as colored lines.}\label{fig3}
\end{figure}

\begin{figure}
\onefigure[width=\columnwidth]{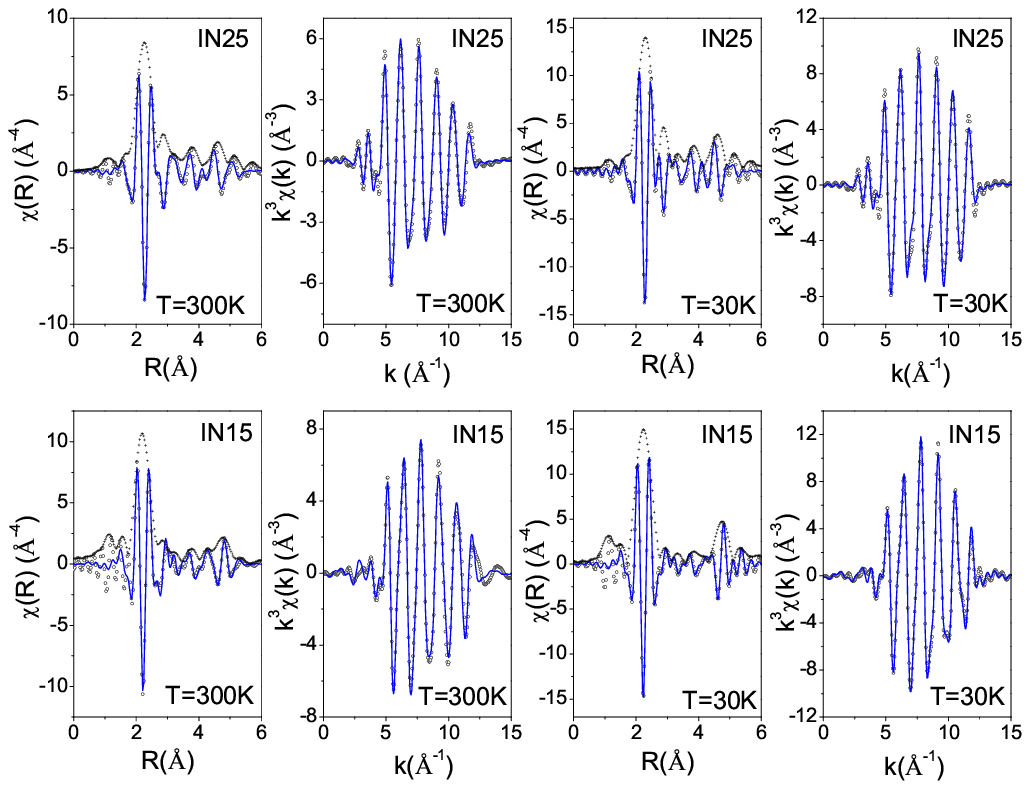}
  \caption{FT Magnitude and real component of EXAFS spectra in R space and the real component in the back transformed $k$ space for
Mn K-edge in IN25 and IN15 at 300K and 30K. The fitting to the data are shown as colored lines.}\label{fig4}
\end{figure}

In order to extract structural parameters like bond distances and $\sigma^2$  EXAFS spectra at Ni and Mn K-edges were fitted using the respective structural models. In case of  IN25 spectra at all temperatures were fitted using correlations based on L2$_1$ structure (cubic model) while the EXAFS spectra in case of IN15, barring those at 300K were fitted using correlations based on the orthorhombic structure described in Ref. \cite{brown}. The 300K EXAFS spectra in IN15 were fitted using the cubic model. Further in case of IN15, the coordination spheres involving Mn/In atoms as backscatters at the same bond distance were fitted as two separate correlations consisting entirely of Mn and In atoms with the coordination number fixed as per their composition ratio. The fitting to the EXAFS data at two representative temperatures, 300K and 30K at the Ni and Mn K edges are shown in Figures \ref{fig3} and \ref{fig4} respectively. 

\begin{figure}
\onefigure[width=\columnwidth]{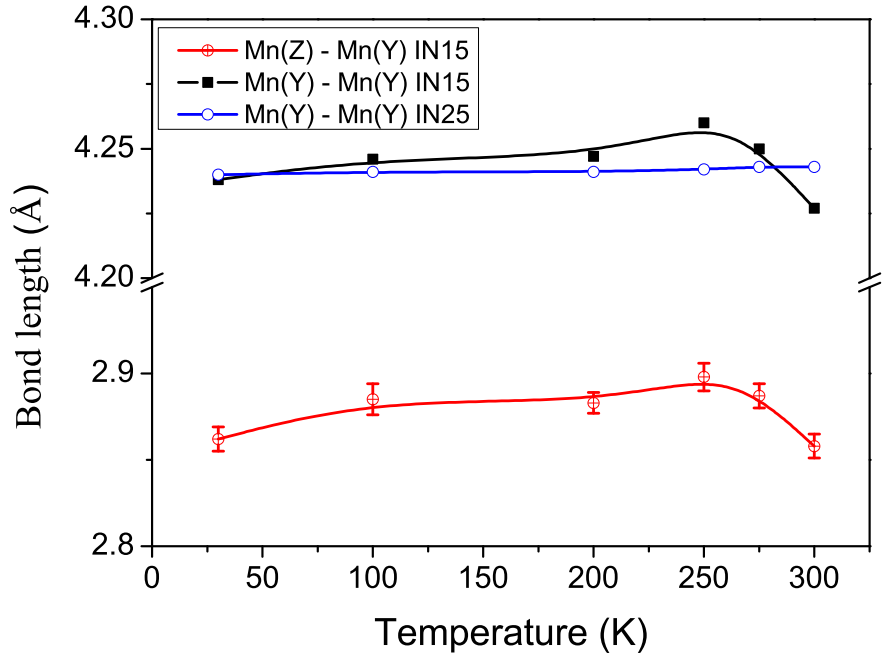}
  \caption{Variation of different Mn-Mn bond lengths as a function of temperature in IN25 and IN15. The vertical line marked as
  T$_M$ is the martensitic temperature of IN15.}
  \label{fig5}
\end{figure}

\begin{figure}
\onefigure[width=\columnwidth]{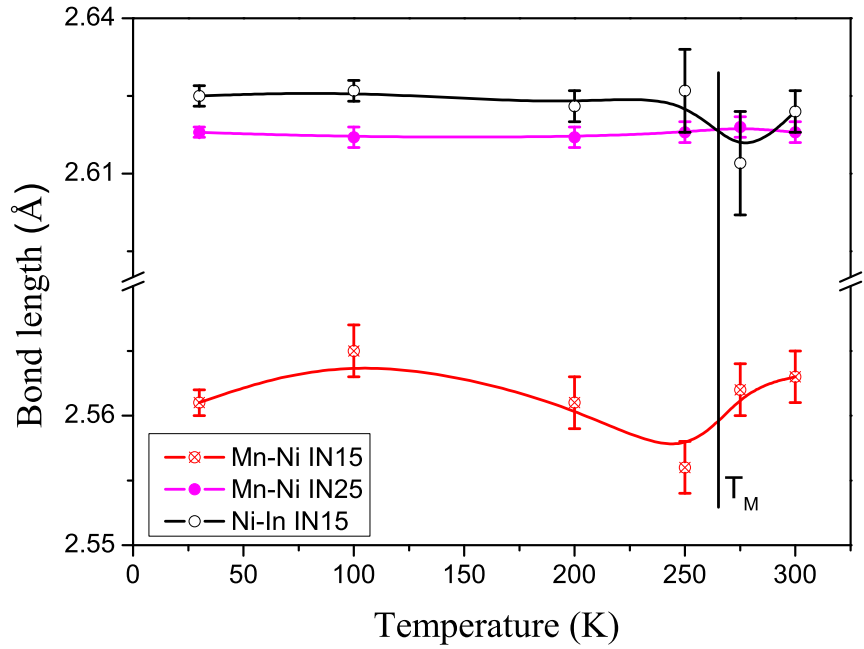}
  \caption{Temperature dependent variation of Mn-Ni and  Ni-In bond lengths in IN25 and IN15. The vertical line marked as
  T$_M$ is the martensitic temperature of IN15.}
  \label{fig6}
\end{figure}

In the X$_2$YZ type Heusler structure, X(Ni) atom is equidistant from Y(Mn) and Z(In). This is indeed true in case of IN25. The Ni-Mn and Ni-In bond lengths obtained from Ni EXAFS at 300K are equal at 2.62\AA \ and in good agreement with the bond distances calculated from lattice constant obtained from XRD. There is also a good agreement of other bond distances obtained from EXAFS analysis with those calculated from lattice constant. The values of bond lengths (R) and the corresponding mean square radia distortion ($\sigma^2$) obtained from fitting at 300K are given in Table \ref{In25} In case of IN15 however, the Ni-Mn and Ni-In distances are different at 2.55\AA \ and 2.62\AA \ respectively even at 300K (see Table \ref{In25}). A difference in these bond distances in the austenitic phase indicates a presence of local structural disorder which has been shown to be due to the presence of excess Mn in the alloy \cite{nelson}. The substituted Mn at the Z site being smaller in size compared to In, is displaced from its position giving rise to a local structural distortion. Therefore a shorter Ni-Mn distance implies Mn(Z) is closer to Ni than Mn(Y) or In. Such a disorder in the nearest neighbor bond distances of Ni was also seen in Ni$_{50}$Mn$_{35}$Sn$_{15}$ in its austenitic phase \cite{pab3}. Furthermore, in IN15, there is an Mn-Mn bond distance at about 3\AA \ between Mn(Y) and Mn(Z) atoms. The bond distance between the two Mn atoms occupying the same sublattice is 4.2\AA. Given the fact that ferromagnetism in both the alloys is due to RKKY interaction between Mn atoms in their own sublattice while the antiferromagnetism in IN15 could be either due to RKKY interaction between Mn(Y) and Mn(Z) or due to Ni-Mn hybridization \cite{buch2,sasi2} it is prudent to examine the temperature variation of Ni-Mn and Mn-Mn bond distances in IN15  and compare them with similar ones in IN25 and Ni-In in IN15. Such a comparison is provided in Figures \ref{fig5} and \ref{fig6}.

\begin{table}
\caption{\label{In25} Results of the fits to the Ni and Mn edge EXAFS data for IN25 and IN15 at 300K. R refers to the bond length and $\sigma^2$ is the thermal mean-square variation in the bond length. The fittings were carried out in the k range: 2 -- 15 \AA$^{-1}$ with k weight = 3 and R range 1-- 3 \AA \ for Ni K edge and 1 -- 5 \AA \ for Mn K edge. Figures in parentheses indicate uncertainty in the last digit.}
\centering
\begin{tabular}{llllll}
\hline
 Atom and &\multicolumn{2}{c}{IN25} & Atom  and & \multicolumn{2}{c}{IN15} \\
Coord.No. & R (\AA) & $\sigma^2$ (\AA$^2$) & Coord. No. & R (\AA) & $\sigma^2$ (\AA$^2$)\\
\hline
 \multicolumn{6}{c}{Ni K edge}  \\
Mn $\times$ 4& 2.621(3) & 0.013(1) & Mn $\times$ 5.44& 2.563(4) & 0.014(1) \\
In $\times$ 4& 2.621(3) & 0.009(1) & In $\times$ 2.56& 2.622(3) & 0.006(1)  \\
Ni $\times$ 12& 3.027(3) & 0.036(5) & Ni $\times$ 12 & 3.03(3) & 0.029(4) \\
 \multicolumn{6}{c}{Mn K edge}  \\
Ni $\times$ 8& 2.618(2) & 0.010(1) & Ni $\times$ 8& 2.563(2) & 0.009(1) \\
In $\times$ 6& 3.020 & 0.017(2)  & In $\times$ 3.8& 2.836(7) & 0.007(1) \\
&&& Mn $\times$ 2.2& 2.858(7) & 0.004(1)\\
Mn $\times$ 12& 4.26(1) & 0.018(2)& Mn $\times$ 12 & 4.22(6) & 0.028(4) \\
Ni $\times$ 8& 4.99(1) & 0.022(4) & Ni $\times$ 8& 4.99(1) & 0.022(4) \\
\hline
\end{tabular}
\end{table}

IN25 is a ferromagnet with T$_c$ = 306K and has a stable crystal structure, thus there is very little or no
variation in the bond distance between Mn atoms Mn(Y)-Mn(Y) in the temperature range 300K to 30K. On the other
hand, the Mn(Y)-Mn(Y) bond distance in IN15 increases in transformation region (300K - 200K) and then decreases
at lower temperatures. The increase in Mn(Y)-Mn(Y) bond distance can be related to the  weakning of ferromagnetism due to martensitic transformation. Therefore any decrease in this bond length at lower temperature should lead to strengthening of ferromagnetism. This is indeed the case. Mn(Y)-Mn(Y) bond distance decreases below 200K and the magentization curve too shows a build up of ferromagnetism. Magnetic hysteresis studies reported on similar composition is also in agreement with this hypothesis \cite{pab2}. Thus variation of bond distances obtained from EXAFS can explain the evolution of ferromagnetic interactions quite well. As regards to the origin of antiferromagnetic interactions, it can be seen from Figure \ref{fig5} that the behavior of Mn(Y)-Mn(Z) bond distance is similar to that of Mn(Y)-Mn(Y) indicating an enhancement of strength of antiferromagnetic interactions at low temperatures. Does this mean antiferromagnetism is also due to RKKY
interactions? Is there any role at all of the nearest neighbour interaction in magnetism of martensitic phase?

To seek answers to these questions, variation of Ni-Mn bond distance as a function of temperature is plotted and
compared with Ni-In bond length in IN15 and Ni-Mn/In bond distance in IN25 in Figure \ref{fig6}. Expectedly, in
case of IN25, the equal Ni-Mn and Ni-In bond distances show almost no variation as a function of temperature. On the other hand, due to structural instability in IN15 the Ni-In bond length in shows a change near T$_M$ and then remains fairly constant at lower temperature. If
Ni-Mn nearest neighbour correlation had no role to play in the magnetism of these alloys its variation should be
similar to Ni-In bond length. However, the variation of Ni-Mn bond distance in IN15 is not only different from
that of Ni-In and Mn-Mn bond lengths, it is also very interesting. Firstly, the variation cannot be explained on the basis of variation of Mn(Y)-Mn(Z) bond distance as the two are distinctly different from each other. If the two were related then the temperature dependence of both these bond distances should have been similar. Secondly, the variation of Ni-Mn quite closely replicates the behaviour of ZFC magnetization curve. Ni-Mn bond distance first decreases quite sharply in the martensitic transformation
region and then increases below 250K and culminating into a broad hump centred around 100K. A shorter Ni-Mn bond
distance implies a stronger hybridization between these two atoms. This will lead to a charge transfer from
nearly full Ni $3d$ band to half filled Mn $3d$ band forcing the latter to be closer to $E_F$. Coupled with
destabilization of structural order due to martensitic transformation will result in complete or partial destruction of ferromagentism. This explains the observed large decrease in magnetization seen at T$_M$. The increase in Ni-Mn bond distance below 250K results in strengthening
of magnetism as evidenced from reordering of Mn spins and a magnetic transition in the martensitic state at
T$_c^M$. Below 100K, the Ni-Mn bond distance shows a tendency to decrease which results in strengthening of
antiferromagnetic interactions. The observed exchange bias effect at lower temperatures in these alloys supports
this argument \cite{pab2}. These facts clearly bring out the importance of hybridization between nearest neighbors in the
magnetism of martensitic state. While it seems from the EXAFS studies Mn(Y)-Mn(Y) interactions are solely responsible
for ferromagnetic interactions, the antiferromagnetic interactions are a result of increase in the hybridization between Ni$3d$ states with Mn$3d$ states due to shortening of Ni-Mn bond length.

In summary, EXAFS studies carried out on Ni$_{50}$Mn$_{25+x}$In$_{25-x}$ as a function of temperature have been
used to study the evolution of nearest neighbor Ni-Mn, Ni-In and Mn-Mn bond distances. The temperature dependence
of these bond lengths suggest that the ferromagnetic interactions depend on the distance between Mn
atoms occupying Y sublattice of the X$_2$YZ Heusler structure. It is the hybridization between Ni$3d$ and
Mn$3d$ that plays a crucial role in the strength of the antiferromagnetic interactions in the martensitic phase.

\acknowledgments
The authors would like to thank Photon Factory for beamtime and assistance under
the proposal No. 2009G214. Two of us (KRP and DNL) are grateful to Department of Science and Technology and Prof
A. K. Raychaudhuri for travel funding under Utilization of Synchrotron and Neutron Facilities Programme.
Financial assistance from Council of Scientific and Industrial Research (CSIR) under the scheme No.
03(1100)/2007/EMR-II is also gratefully acknowledged. PAB would like to acknowledge JSPS for the fellowship.

\newpage

\end{document}